\documentstyle[aps,prb,epsf,twocolumn,epsfig,floats]{revtex}

\begin{document}
\draft




\wideabs{

\title{Steady-state and equilibrium vortex configurations,
transitions, and evolution in a mesoscopic superconducting
cylinder} %
\author{Sangbum Kim} %
\address{Department of Mechanical engineering, Texas A\&M
University, College Station, Texas 77843} %
\author{Chia-Ren Hu} %
\address{Center for Theoretical Physics, Department of Physics,
Texas A\&M University, College Station, Texas 77843} %
\author{Malcolm J. Andrews} %
\address{Department of Mechanical engineering, Texas A\&M
University, College Station, Texas 77843} %

\date{January 13, 2003} %
\maketitle %
\begin{abstract} %

A numerical scheme to study the mixed states in a mesoscopic
type-II superconducting cylinder is described. Steady-state
configurations and transient behavior of the magnetic vortices for
various values of the applied magnetic field $H$ are presented.
Transitions between different multi-vortex states as $H$ is
changed is demonstrated by contour plots and jumps in the $B$ vs
$H$ plot. Evolving a uniformly-superconducting initial state using
the simplest set of relaxation equations shows that the system
passes through nearly metastable intermediate configurations,
while seeking the final minimum-energy, steady state consistent
with the square symmetry of the sample. An efficient scheme to
determine the equilibrium vortex configuration in a mesoscopic
system at any given applied field, not limited to the symmetry of
the system, is devised and demonstrated.

\end{abstract}
} 

\section {Introduction}

The electromagnetic characteristics of type-II superconductors
have been of great interest, particularly since the discovery of
high-temperature superconductors. Furthermore, a recent surge in
interest about nano-technology, has drawn the attention of
researchers from various disciplines to a detailed understanding
of the characteristics of mesoscopic superconducting samples.
Several phenomenological theories have been developed during
decades of superconductor research, one popular choice is the
Ginzburg-Landau theory.~\cite{deGennes,Chapman} Its time-dependent
extension is known as Time-Dependent Ginzburg-Landau (TDGL)
theory.~\cite{Tinkham}

In the Ginzburg-Landau theory, the electromagnetic state of a
superconductor can be determined by solving a system of partial
differential equations. It was discovered by Abrikosov that if the
$\kappa$ parameter that appears in the equations, now known as the
Ginzburg-Landau parameter, is larger than $1/\sqrt{2}$, then when
a bulk superconductor is placed in a sufficiently large magnetic
field, the magnetic field penetrates the superconductor in the
form of singly-quantized vortices. Around each vortex flows a
supercurrent,~\cite{Abrikosov} confining a single quantum of
magnetic flux within it. Superconductors with this property are
known as type-II.

Here we present the result of a numerical study about the
magnetization process inside a superconducting cylinder with
sub-micron lateral dimension in an external magnetic field. We
have restricted the work to a square cross section of a linear
size equal to 4.65 times $\lambda$ (the magnetic penetration
depth). A typical value for $\lambda$ is about 500 $\AA$, and then
the cross-sectional area is about 0.054 $\mu m^{2}$.

Previous works on the magnetization of a mesoscopic superconductor
without pinning centers have been reported by Peeters {\em et al.}
~\cite{BaelPeet,SchPeet,SchPeetDeo,DeoSchPeet} and
others~\cite{Bonca,Misko-etal-02}, who have presented extensive
calculations on the superconducting state in mesoscopic, type-I,
superconducting thin films. In most cases they found transitions
between giant vortex states of different circulation quantum
numbers $L$, with some multi-vortex states occasionally appearing
as thermodynamically stable states, but mostly as metastable
states. These predictions appear to have already received some
level of experimental confirmation, although some discrepancies
still exist~\cite{Geim-etal}. (Ref.~\cite{Misko-etal-02} have
mainly compared the energy of a "3-2" vortex-antivortex molecule
state with that of a single off-centered vortex state at $L = 1$,
as both evolve to the equilibrium state of a single vortex at the
center.) Misko et al.~\cite{Misko-etal-03} have studied both
type-I and type-II mesoscopic trianglular cylinders, and have
shown that a vortex-antivortex molecule appears only if the sample
is type-I. They considered only one field value at which $L = 2$
is favored, and did not consider vortex configurational
transitions as the field changes.~\cite{note3}

Our aim is to simulate how vortices enter and settle in stable
arrangements when a mesoscopic type-II superconductor of a given
symmetry is first cooled below the critical temperature, and then
an external magnetic field is applied. This is often termed zero
field cooling (ZFC). We find that only vortex numbers and
configurations consistent with the sample symmetry can appear in
this case. It is known that there exist global minimum-energy
vortex configurations with reduced symmetry, correspondingly to
final equilibrium states at general values of the applied field.
To find these equilibrium states we developed, and demonstrate
here, an efficient numerical scheme.

Our approach is to solve a set of simplified (and discretized)
TDGL equations, in which the coupling to the electric field is
neglected, and the superconducting order parameter and the
magnetic field are assumed to relax with the same time scale.
These assumptions are not physical, but are acceptable here, since
we are only interested in obtaining the final steady-state vortex
configurations, and the symmetry-related qualitative behavior of
the transient configurations and their evolution. (For a more
physical set of TDGL equations see Tinkham~\cite{Tinkham}. For an
example of numerical solution of such a set of TDGL equations,
see~\cite{Crabtree-etal}.) However, since the equations we have
solved do not contain any thermal fluctuation terms, and the
sample we considered has a perfect square symmetry, we find that
when starting with the Meissner state with no field penetration,
then the final steady-state vortex configurations we obtain all
have perfect square symmetry, with vortex numbers also limited to
only multiples of four (the symmetry number). These configurations
would correspond to physical situations under zero-field cooling,
if the physical sample has indeed perfect symmetry, and the
temperature is sufficiently low, so that thermal fluctuations are
not able to overcome any energy barrier for vortex entry,
expulsion, or rearrangement. However, if the sample surface has
slight imperfection, or if the temperature is not sufficiently
low, then these configurations are, in most cases, not in
equilibrium at the given magnetic field strength. Even the vortex
number may not be correct. However, if we insert into the
equations terms to simulate thermal fluctuation, as in the method
of simulated annealing,~\cite{SimuAnneal,Doria} then the
simulation program will take a much longer time to run, and may
become impractical even with a supercomputer. So we have devised
an efficient scheme to find the equilibrium vortex configurations:
We solve the same set of relaxation equations without any thermal
fluctuation terms. But instead of starting the solution with the
Meissner state as the initial state, we devise artificial initial
states with a given number of vortices in random positions. We
present analytic expressions for such initial states, in terms of
a widely known approximate expression for a singly-quantized
vortex in cylinder coordinates. Then for vortex numbers not too
different from the equilibrium number, the final steady states
obtained by solving our relaxation equations will, in most cases,
have the number of vortices close to those of the initial states.
By comparing the total Gibbs energies of these steady states with
different vortex numbers --- sometimes we obtain more than one
configuration for the same vortex number, (when the vortex number
exceeds four,) then their Gibbs energies are also compared
--- we can find the state with the lowest total Gibbs
energy, which we identify as the equilibrium state with the
equilibrium vortex number. We give an explicit demonstration of
this scheme~\cite{note1}, that might be very useful in view of the
recent interest in nanoscience and nanotechnology. In some
nano-applications one may need to find the equilibrium vortex
configurations in a mesoscopic superconducting sample, at
different applied magnetic fields. We note that in a bulk sample
vortices like to form a triangular lattice. Thus, when the sample
does not conform with this symmetry, and if the sample is
sufficiently small so that the boundary effect on the equilibrium
vortex configurations is important, then the system is frustrated,
and the equilibrium vortex configurations can be quite intriguing,
and difficult to foresee. The scheme devised here is then very
convenient and useful for finding the answer.~\cite{note2}

\section {Time-dependent Ginzburg-Landau model}

In an external magnetic field {\bf H}, the Gibbs free energy
density $g$ of a superconducting state is given by~\cite{Tinkham}
\begin{eqnarray}
g & = & f_{n}+\alpha |\Psi|^{2}+{\beta\over 2}|\Psi|^{4} \nonumber \\
  &   & \mbox{}+{1\over
2m_{s}}\left|\left(-\imath\hbar\nabla-{e_{s}\over c}{\bf A}
    \right)\Psi\right|^{2}         +{|{\bf h}|^2\over
{8\pi}}-{{\bf h}\cdot{\bf H}\over{4\pi}} \end{eqnarray} where
$f_{n}$ is the free energy density in the normal state in the
absence of the field, $\Psi$ is the complex-valued order
parameter, with the superscript * denoting complex conjugation,
{\bf A} the magnetic vector potential, ${\bf h}=\nabla\times {\bf
A}$ the induced magnetic field and {\bf H} the applied magnetic
field. The supercurrent density is expressed as ${\bf
J}_{s}=\nabla\times\nabla\times {\bf A} ={{e_{s} \hbar}\over
{2\imath m_{s}}}\left( \Psi^{*}\nabla\Psi-\Psi\nabla\Psi^{*}
\right) -{e_{s}^{2}\over {m_{s} c}} \left| \Psi \right|^{2} {\bf
A}$. In the above $e_{s}$ is the "effective charge" of a Cooper
pair which is twice the charge of an electron, and $m_{s}$ its
"effective mass" which can be selected arbitrarily, but the
conventional choice is twice the mass of an electron. Also, $c$ is
the speed of light, and $\hbar=h/2\pi$ where $h$ is Planck's
constant.

Ginzburg-Landau theory postulates that the Gibbs free energy of a
superconducting sample $\Omega$, $G\left(\Psi,{\bf A}\right) =
\int_{\Omega} g d\Omega$ is at a minimum in the superconducting
state. The celebrated Ginzburg-Landau equations are obtained by
minimizing this functional with respect to $\Psi$ and ${\bf A}$
using the variational principle.

Since a constant term does not change the end result of the
variational technique, an algebraic manipulation is made to
subtract $f_{n}$ and add ${\bf H}\cdot{\bf H}/8\pi$ to the $g$
above, giving~\cite{Du}
\begin{eqnarray}
G\left(\Psi,{\bf A}\right) & = & \int_{\Omega} \left( \alpha
|\Psi|^{2}+{\beta\over 2}|\Psi|^{4} + {|{\bf h}-{\bf H}|^2\over
{8\pi}} \right. \nonumber \\ %
& & \left. \mbox{}+{1\over 2m_{s}}\left|\left(-\imath\hbar\nabla
- {e_{s}\over c}{\bf A} \right)\Psi\right|^{2} \right) d\Omega %
\end{eqnarray}

Nondimensionalizing variables as

$x^{\prime}={x\over\lambda}, {\bf H}^{\prime}={{\bf H}\over
{\sqrt{2} H_{c}}}$, ${\bf h}^{\prime}={{\bf h}\over {\sqrt{2}
H_{c}}}$, ${\bf j}^{\prime}={{2\sqrt{2}\pi\lambda}\over {c H_{c}}}
{\bf j}$, ${\bf A}^{\prime}={{\bf A}\over {\sqrt{2} H_{c}
\lambda}}$, $\Psi^{\prime}={\Psi\over \Psi_{0}}\,.$
The characteristic scales are:
$\left|\Psi_{0}\right|=\sqrt{-\alpha/\beta}$ which is the
magnitude of $\Psi$ that minimizes the free energy in the absence
of a field; the thermodynamic critical field strength
$H_{c}=\left( 4\pi|\alpha|\left|\Psi_{0}\right|^{2}
\right)^{1/2}$, which divides the normal state and superconducting
state regions in Type-I superconductor phase diagram; the London
penetration depth $\lambda=\left({{m_{s} c^{2}} \over
{4\pi\left|\Psi_{0}\right|^{2} e_{s}^{2}}} \right)^{1/2}$; the
coherence length $\xi=\left( {\hbar^{2} \over {2m_{s} |\alpha|}}
\right)^{1/2}$; and, the Ginzburg-Landau parameter
$\kappa=\lambda/\xi$.

We obtain the dimensionless gauge-invariant free energy functional
(omitting primes for convenience),
\begin{eqnarray}
G\left(\Psi,{\bf A}\right) & = & \int_{\Omega} \left(
-|\Psi|^{2}+{1\over 2}|\Psi|^{4}+|\nabla\times {\bf A}-{\bf H}|^{2} \right. \nonumber \\
   &   & \left. \mbox{}+\left|\left({\nabla\over\kappa}
         -\imath {\bf A} \right)\Psi\right|^{2} \right) d\Omega
\end{eqnarray}

The simplified TDGL model we employ to find solutions of the
static GL equations may be viewed as a gradient flow with the
energy functional. That is, the variation of $\left( \Psi, {\bf A}
\right)$ w.r.t. time should be in the opposite direction of the
gradient of the energy functional, ${{\partial\Psi}\over {\partial
t}}=-{{\partial G}\over {\partial\Psi^{*}}}$, ${{\partial {\bf
A}}\over {\partial t}}=-{1\over 2}{{\partial G}\over {\partial
{\bf A}}}$ with time, $t$, in units of the only relaxation time of
the equations. The natural boundary conditions are: (1) the
continuity of the parallel component of the magnetic field across
the boundary surface:
$\left( \nabla\times {\bf A} \right)\times {\bf n} = {\bf
H}\times{\bf n}$, (for two-dimensional problems only, see below)
and (2) the vanishing gauge-invariant normal derivative of $\Psi$:
$\left( {\nabla\over\kappa}-\imath {\bf A} \right)\Psi\cdot {\bf
n}=0$,~\cite{Coskun} with ${\bf n}$ denoting the outward surface
normal.

\section {Discretization and calculation procedure}

For long cylindrical samples, we need only solve a 2-D problem. We
take ${\bf A}=\left( A(x,y),B(x,y),0 \right)$ and ${\bf
H}=(0,0,H)$ where $H=\left( \nabla\times {\bf A}
\right)_{z}={{\partial B}\over {\partial x}}-{{\partial A}\over
{\partial y}}$.

Defining the link variables as below, the gauge invariance is
preserved in discretizing the Gibbs free energy and the consequent
TDGL equations. $W(x,y) = exp\left(\imath\kappa\int^{x}
A(\varsigma, y)d\varsigma\right)$, $V(x,y) =
exp\left(\imath\kappa\int^{y} B(x,\eta)d\eta\right)$
Noting that
$\left|\partial_{x}\left(W^{*}\Psi\right)\right|=\left|\left(
\partial_{x}-\imath\kappa A\right)\Psi\right|$, and
$\left|\partial_{y}\left(V^{*}\Psi\right)\right|=\left|\left(
\partial_{y}-\imath\kappa B\right)\Psi\right|$, we have
\begin{eqnarray}
G\left(\Psi,{\bf A}\right) & = & \int_{\Omega} \left(
-|\Psi|^{2}+{1\over 2}|\Psi|^{4} + |\nabla\times {\bf A}-{\bf
H}|^{2} \right. \nonumber \\
& & \left. \mbox{}+\left|{1\over\kappa}\partial_{x}\left(
   W^{*}\Psi\right)\right|^{2}
         +\left|{1\over\kappa}\partial_{y}\left(
   V^{*}\Psi\right)\right|^{2} \right) d\Omega\,.
\end{eqnarray}

\begin{figure}[htb]
\begin{center}
\begin{picture}(120,120)
\put (5,20){\line(1,0){105}} \put (5,60){\line(1,0){105}} \put
(5,100){\line(1,0){105}}

\put (20,5){\line(0,1){105}} \put (60,5){\line(0,1){105}} \put
(100,5){\line(0,1){105}}

\put (10,40){\line(1,0){5}} \put (20,40){\line(1,0){5}} \put
(30,40){\line(1,0){5}} \put (40,40){\line(1,0){5}} \put
(50,40){\line(1,0){5}} \put (60,40){\line(1,0){5}} \put
(70,40){\line(1,0){5}} \put (80,40){\line(1,0){5}} \put
(90,40){\line(1,0){5}} \put (100,40){\line(1,0){5}} \put
(110,40){\line(1,0){5}}

\put (10,80){\line(1,0){5}} \put (20,80){\line(1,0){5}} \put
(30,80){\line(1,0){5}} \put (40,80){\line(1,0){5}} \put
(50,80){\line(1,0){5}} \put (60,80){\line(1,0){5}} \put
(70,80){\line(1,0){5}} \put (80,80){\line(1,0){5}} \put
(90,80){\line(1,0){5}} \put (100,80){\line(1,0){5}} \put
(110,80){\line(1,0){5}}

\put (40,10){\line(0,1){5}} \put (40,20){\line(0,1){5}} \put
(40,30){\line(0,1){5}} \put (40,40){\line(0,1){5}} \put
(40,50){\line(0,1){5}} \put (40,60){\line(0,1){5}} \put
(40,70){\line(0,1){5}} \put (40,80){\line(0,1){5}} \put
(40,90){\line(0,1){5}} \put (40,100){\line(0,1){5}} \put
(40,110){\line(0,1){5}}

\put (80,10){\line(0,1){5}} \put (80,20){\line(0,1){5}} \put
(80,30){\line(0,1){5}} \put (80,40){\line(0,1){5}} \put
(80,50){\line(0,1){5}} \put (80,60){\line(0,1){5}} \put
(80,70){\line(0,1){5}} \put (80,80){\line(0,1){5}} \put
(80,90){\line(0,1){5}} \put (80,100){\line(0,1){5}} \put
(80,110){\line(0,1){5}}

\put (62,62){P} \put (22,62){W} \put (102,62){E} \put (62,22){S}
\put (62,102){N}

\put (42,62){w} \put (82,62){e} \put (62,42){s} \put (62,82){n}
\end{picture}
\end{center}
\caption{Staggered Grid}
\end{figure}
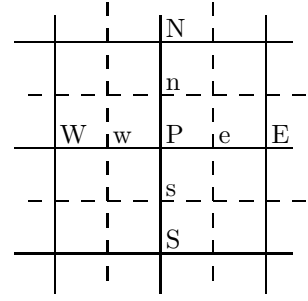

We discretize the free energy functional on a staggered grid in
$\Omega$ shown in Fig. 1.~\cite{Adler,Kaper} This gives us a
second-order approximation in $h_{x}$ and $h_{y}$ to the
continuous energy functional, where the $h_{x}$ and $h_{y}$ are
the spatial increments in the x- and y-direction. The staggered
grid also leads to a satisfactory way of discretizing the natural
boundary conditions.~\cite{Wang} For a rectangular grid, the first
component of the vector potential is constant in time on one pair
of the edges of the boundary, and the second component is constant
in time on the other pair.~\cite{Coskun}

In this paper, we assume that the cylindrical superconductor has a
square cross-section and is subject to an applied field along the
cylindrical axis. The applied field is assumed to be constant in
time. We further assume the order parameter $\Psi$ varies in the
cross-sectional plane of the cylindrical sample, and the vector
potential {\bf A} has only two nonzero components $(A, B)$, which
also lie in this plane. We also assume that the superconductor has
no pinning sites. Then at steady-state conditions, the vortices
settle in maximal distances due to the repulsion of each other.
This requirement leads to triangular lattice of vortices in an
infinitely large domain.~\cite{Tinkham}

In the staggered grid the lattice points for $\Psi$, $A$, and $B$
are all different, with $\Psi$ evaluated at the node center (i,j),
$A$ evaluated at the east cell face (i+1/2, j), and $B$ evaluated
at the north cell face (i,j+1/2). According to
Refs.~\onlinecite{Adler,Kaper}, this formulation keeps second
order accuracy in the derivative evaluations as they appear in
each of the discretized equations.

Now the discrete TDGL is obtained by minimizing the discrete
energy functional $G_d$ with respect to the variation in $\Psi$
and {\bf A} as %
\begin{eqnarray}
{{\partial\Psi_{P}}\over {\partial t}} & = &
{{h_{x}h_{y}}\over\kappa^{2}} \left( {e^{\imath A_{w}\kappa
h_{x}}\Psi_{W}-2\Psi_{P}+{e^{-\imath A_{e}\kappa
h_{x}}\Psi_{E}}\over h_{x}^{2}}+ \right. \nonumber \\ & & \left.
{{e^{\imath B_{s}\kappa h_{y}}\Psi_{S}-2\Psi_{P}+{e^{-\imath
B_{n}\kappa
h_{y}}\Psi_{N}}\over h_{y}^{2}}} \right)+ \nonumber \\
   &   & {h_{x}h_{y}}N_{1}\left(\Psi_{P}\right) \\
{{\partial A_{e}}\over {\partial t}} & = & \mbox{}-h_{x} \left(
{{B_{nE}-B_{n}+B_{s}-B_{sE}}\over h_{x}}- \right. \nonumber \\
& & \left. {{A_{Ne}-2 A_{e}+A_{Se}}\over h_{y}} \right)+ \nonumber \\
   &   & {h_{y}\over\kappa}N_{2}\left(A_{e},\Psi_{P},\Psi_{E}\right) \\
{{\partial B_{n}}\over {\partial t}} & = & \mbox{}-h_{y} \left(
{{A_{w}-A_{e}+A_{Ne}-A_{Nw}}\over h_{y}}- \right. \nonumber \\
& & \left. {{B_{nE}-2 B_{n}+B_{nW}}\over h_{x}} \right)+ \nonumber \\
   &   & {h_{x}\over\kappa}N_{3}\left(B_{n},\Psi_{P},\Psi_{N}\right)
\end{eqnarray} %
where %
\begin{equation} %
N_{1}(\Psi_{P}) = \left(1-|\Psi_{P}|^{2}\right)\Psi_{P} \nonumber
\end{equation} %
\begin{eqnarray} %
N_{2}\left(A_{e},\Psi_{P},\Psi_{E}\right) & = &
\left(\Phi_{P}\Theta_{E}-\Theta_{P}\Phi_{E}\right)\cos
\left(A_{e}\kappa h_{x}\right)- \nonumber \\
   &  &  \left(\Phi_{P}\Phi_{E}+\Theta_{P}\Theta_{E}\right)\sin
         \left(A_{e}\kappa h_{x}\right) \nonumber \end{eqnarray}
\begin{eqnarray} N_{3}\left(B_{n},\Psi_{P},\Psi_{N}\right) & = &
\left(\Phi_{P}\Theta_{N}-\Theta_{P}\Phi_{N}\right)\cos
\left(B_{n}\kappa h_{y}\right)- \nonumber \\   &  &
\left(\Phi_{P}\Phi_{N}+\Theta_{P}\Theta_{N}\right)\sin
\left(B_{n}\kappa h_{y}\right) \nonumber \end{eqnarray} where
$\Theta$ and $\Phi$ are the real and imaginary parts of $\Psi$,
and boundary conditions of the computational domain $\Omega$ (with
$T$, $B$, $L$, and $R$ denoting top, bottom, left, and right,
respectively):
 \begin{eqnarray} \Psi_{P}=\Psi_{S} e^{\imath\kappa h_{y}
B_{s}}, & \hspace{.05in}\textup{on}\hspace{.05in}\Omega_{T} \\
\Psi_{P}=\Psi_{N} e^{-\imath\kappa h_{y} B_{n}}, &
\hspace{.05in}\textup{on}\hspace{.05in}\Omega_{B} \\
\Psi_{P}=\Psi_{E} e^{-\imath\kappa h_{x} A_{e}}, &
\hspace{.05in}\textup{on}\hspace{.05in}\Omega_{L} \\
\Psi_{P}=\Psi_{W} e^{\imath\kappa h_{x} A_{w}}, &
\hspace{.05in}\textup{on}\hspace{.05in}\Omega_{R}
\end{eqnarray}

\begin{eqnarray}
A_{e}=A_{Se}-\left(H-{{B_{nE}-B_{n}}\over h_{x}}\right)h_{y} &
\hspace{.05in}\textup{on}\hspace{.05in}\Omega_{T} \\
A_{e}=A_{Ne}+\left(H-{{B_{nE}-B_{n}}\over h_{x}}\right)h_{y} &
\hspace{.05in}\textup{on}\hspace{.05in}\Omega_{B} \\
B_{n}=B_{nE}-\left(H+{{A_{Ne}-A_{e}}\over h_{y}}\right)h_{x} &
\hspace{.05in}\textup{on}\hspace{.05in}\Omega_{L} \\
B_{n}=B_{nW}+\left(H+{{A_{Ne}-A_{e}}\over h_{y}}\right)h_{x} &
\hspace{.05in}\textup{on}\hspace{.05in}\Omega_{R}
\end{eqnarray}

The finite difference equations are solved by the Euler method
with $h_{x} = h_{y} = 0.15$ and $\Delta t = 0.05$, and taking
$\kappa = 4$. In the numerical computations that follow,
everything is kept the same except for the strength of the applied
magnetic field and/or the initial conditions.

\section {Steady States under zero-field cooling in a perfectly
square sample at low temperatures}

We first solve the above set of equations assuming that the
initial state is the perfect Meissner state with no field
penetration.  As explained in the introduction, this corresponds
to applying a magnetic field after zero-field cooling. Fig. 2
shows the plots of $|\Psi|^{2}$ [the left figure in (a) through
(k)] and $h=\nabla\times {\bf A}$ [the right figure in (a) through
(k)] for the final steady states reached at a sequence of
increasing $H$ values. In the left figures $|\Psi|^{2}$, which is
interpreted physically as the density of Cooper pairs, runs from 0
to 1, the level 1 corresponding to the full superconducting state.
Each isolated group of contours is called a "vortex" representing
the supercurrent J circling around the vortex core, with $\Psi =
0$ at the vortex core. In the 3-D plots of Fig. 2 (b), it is clear
that the vortices reach close to $|\Psi|^{2} = 0$ at the core.

Note that the number of vortices increases in multiples of 4. This
is a consequence of the fact that the vortices are symmetrically
created at the mid-points of the sample edges. Perfect symmetry in
the sample geometry dictates that each side creates an equal
number of vortices. The symmetry manifests itself in this
geometry-dominated problem, and vortices arrange themselves in the
square-symmetric configurations. The resultant steady states are
mostly not true equilibrium states since the vortex arrangements
do not reflect the intrinsic tendency of vortices to form a
triangular lattice known to appear in bulk samples. The natural
next step is to add a thermal fluctuation term to find the true
equilibrium states which may or may not conform with this
symmetry. This approach would then be like simulated
annealing.~\cite{SimuAnneal,Doria} However, we believe it would
not be practical to perfect this approach since it will likely be
difficult to determine the appropriate rate of cooling and
starting temperature. The run-time of the computer program might
also be expected to be much longer than we have found here. So we
have devised a different approach which we believe is much more
efficient for finding the equilibrium states. This is given in a
later section. We shall see that even the cases with low number of
vortices are not the true equilibrium. Also of interest is the
fact that for some H the vortex configuration has a much more
difficult time to settle down, needing much longer run-times to
get to a steady state. Geometry controls the settling time more
than the energy in these cases.

\begin{figure}
\centerline{\epsfig{figure=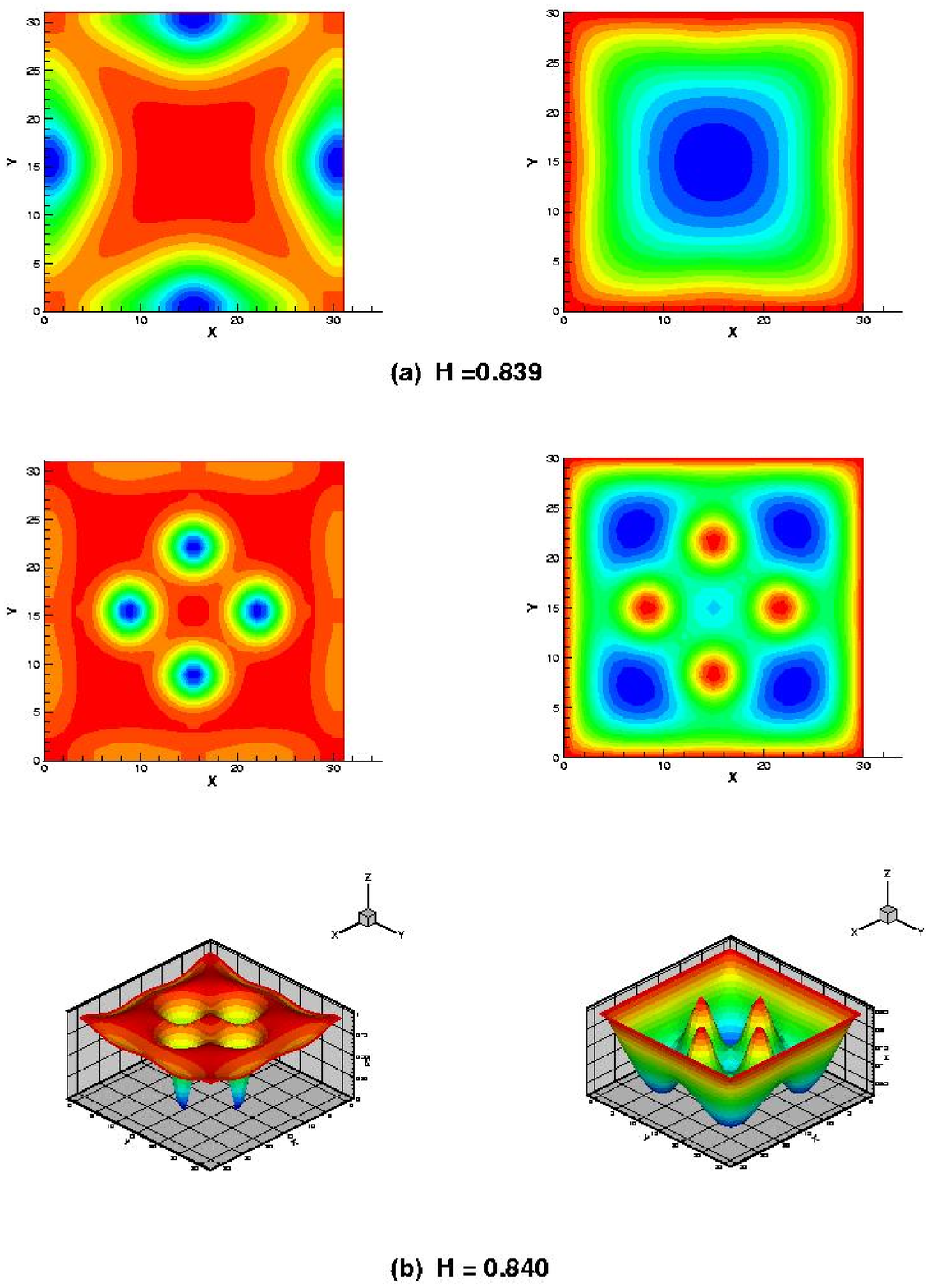,width=8cm}}
\centerline{\epsfig{figure=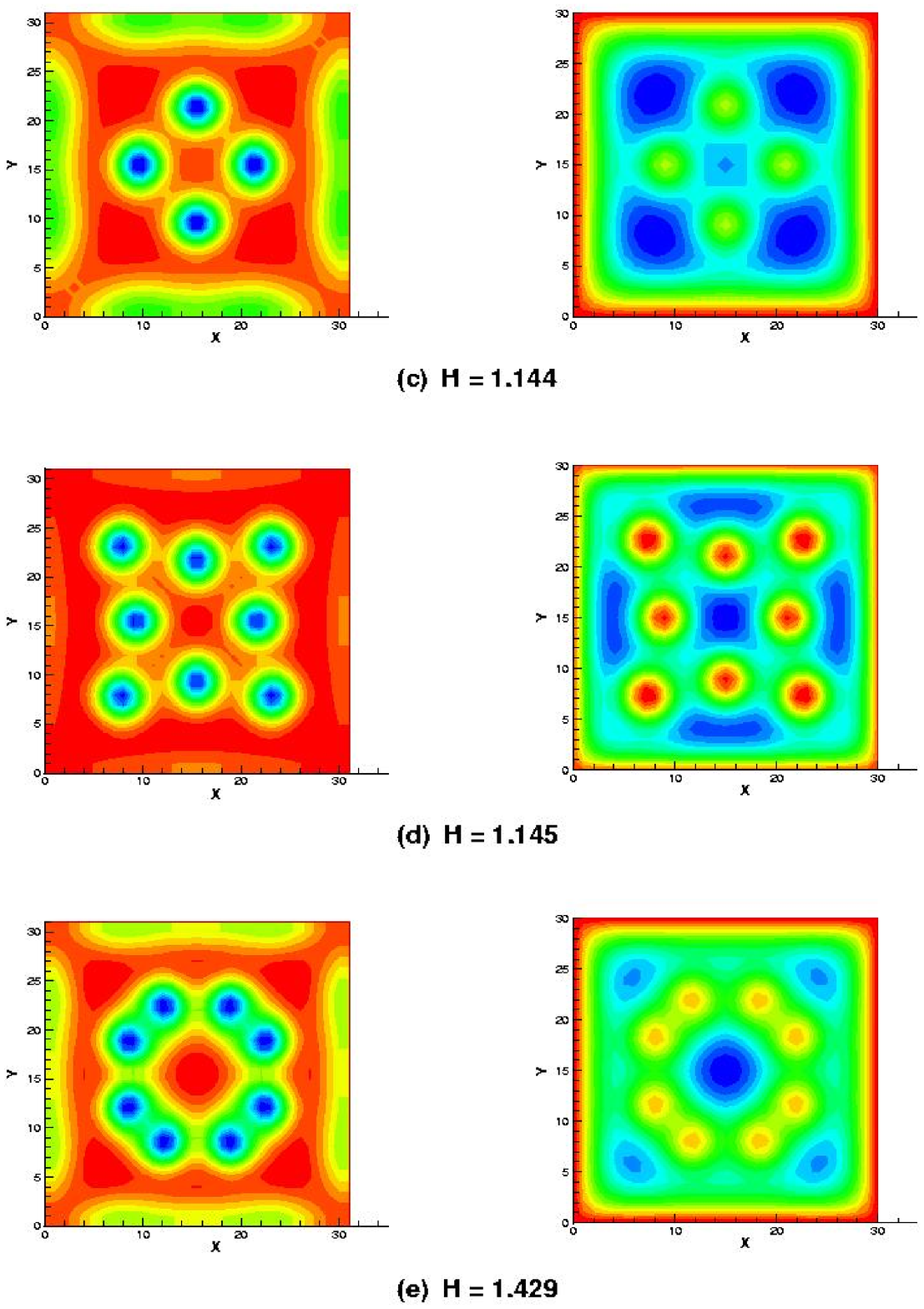,width=8cm}}
\end{figure}
\begin{figure}

\centerline{\epsfig{figure=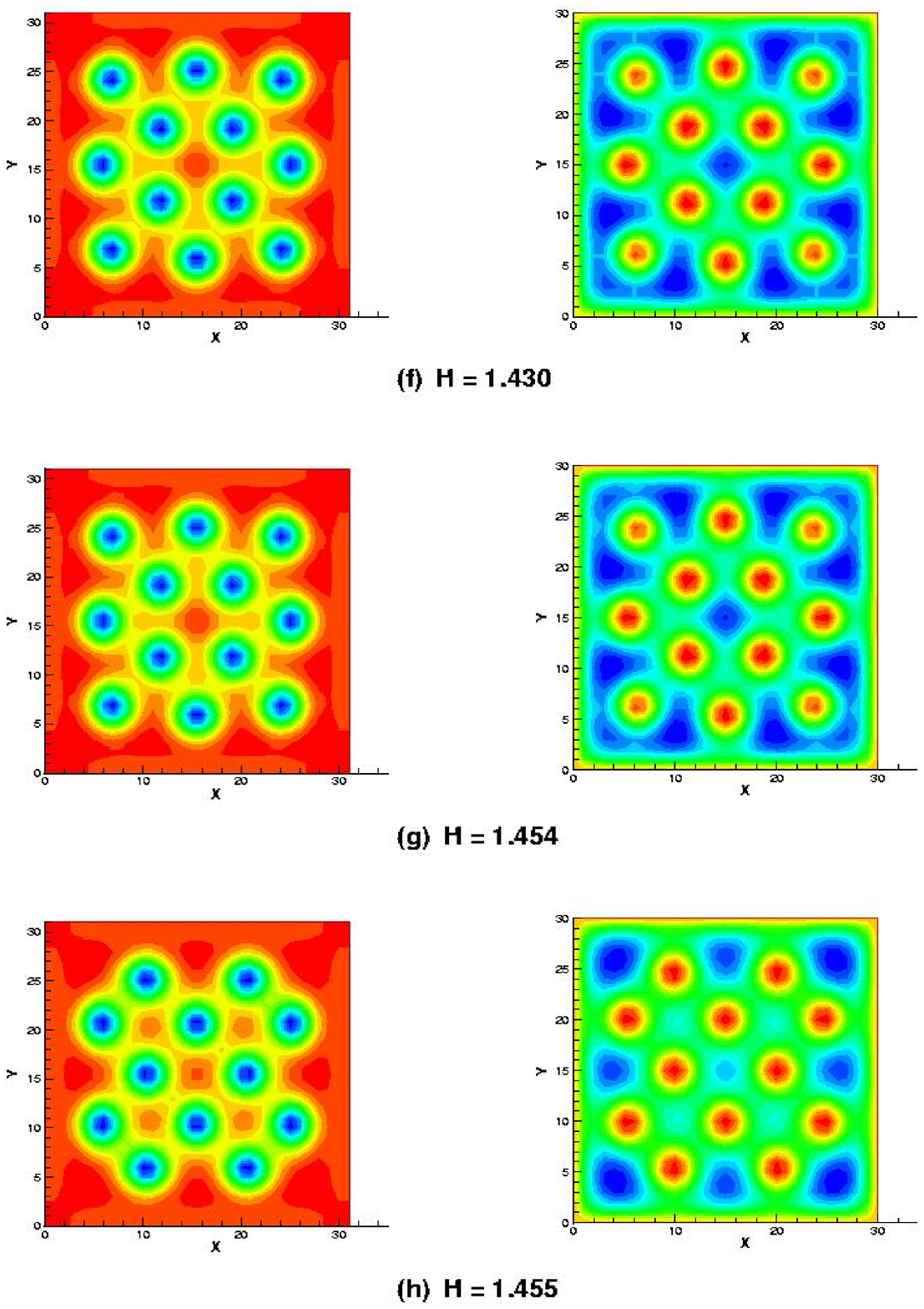,width=8cm}}
\centerline{\epsfig{figure=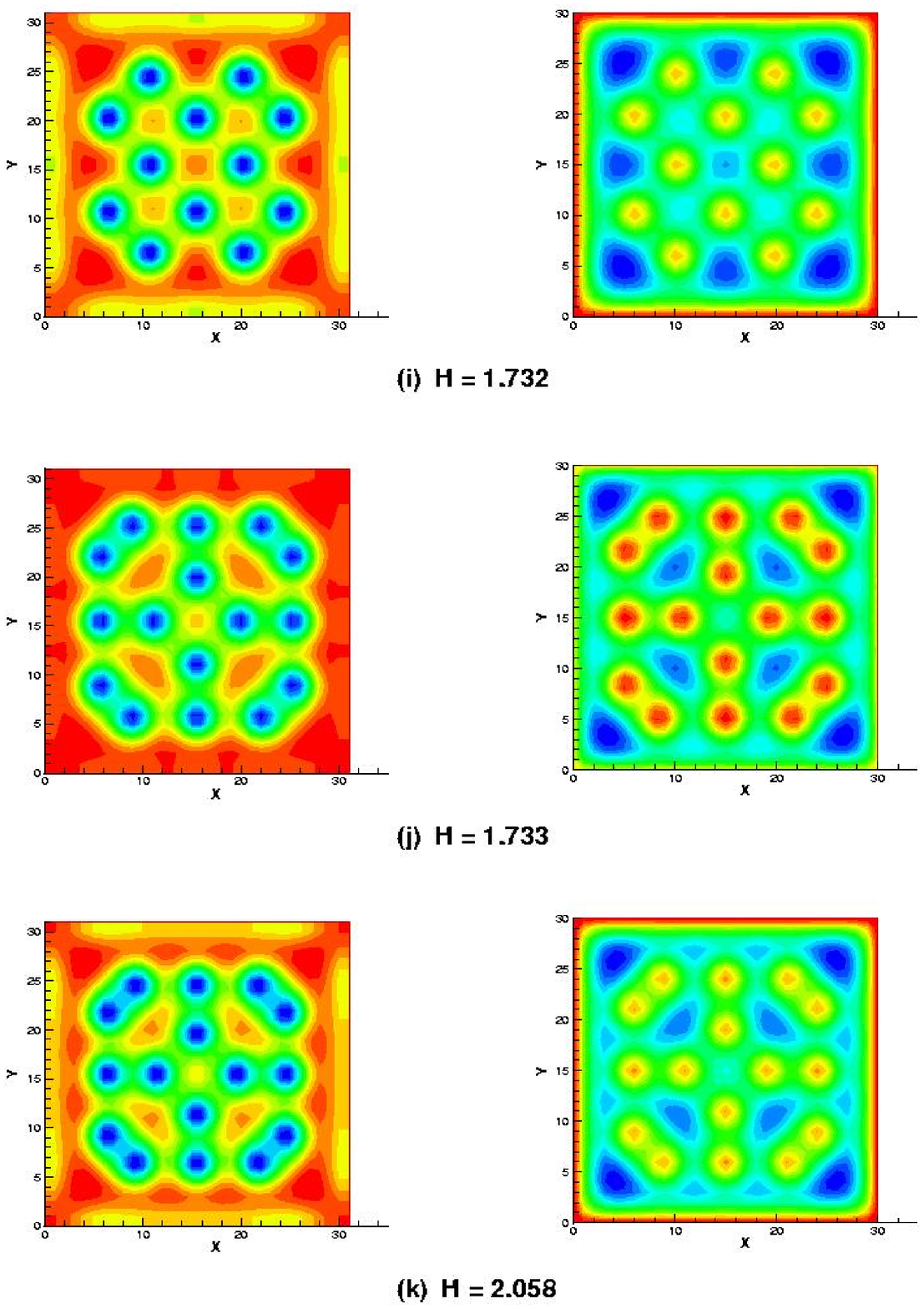,width=8cm}} \vspace{.2cm}
\caption{Plots of $|\Psi|^{2}$ [left figures in (a) through (k)]
and $h = \nabla\times {\bf A}$ [right figures in (a) through (k)]
for various H.} \end{figure}

Our results are summarized in Table 1, which lists the range of
$H$ for each possible number of vortices and the induced magnetic
field $B = {1\over |\Omega|}\int_{\Omega} hd\Omega$. The $H$'s
listed correspond to the threshold values for each range. They
were found on a trial-and-error basis, and can be refined to any
desired accuracy. Below $H = 0.839$, there is no vortex. Between
$H = 0.84$ and 1.144, the vortex number $n = 4$, and so on. Of
course, these thresholds change as we change the sample size. For
example, if we double the length of each side, $H = 1.0$ gives $n
= 36$.

\begin{center}
\begin{tabular}{|c|c|c|}  \hline
n  &  H  &  B \\ \hline
0  &  0.839 &  0.616350 \\ \hline
4  &  0.84  & 0.733198 \\ \hline
4  &  1.144 &  0.977913 \\ \hline
8  &  1.145 &  1.069432 \\ \hline
8  &  1.429 &  1.302241 \\ \hline
12 &  1.43  & 1.379098 \\ \hline
12 &  1.732 & 1.628862 \\ \hline
16 &  1.733 & 1.698629 \\ \hline
16 & 2.058 & 1.971973 \\ \hline
\end{tabular}
\vspace{.2cm}

Table 1. The number of vortices $n$ and the induced magnetic field
$B$ for the applied magnetic field $H$.
\end{center}

The $B$ vs $H$ plot is shown in Fig. 3, and reveals an abrupt
increase in $B$ between the regions of different number of
vortices. For example, when $H$ changes from 1.144 to 1.145, $B$
changes from 0.977913 to 1.069432. If the limit $\Delta H
\rightarrow 0$ is taken, we expect a sudden configurational phase
transition increasing the number of vortices, as is apparent in
Figs. 2 (c) and (d). Such mini-first-order transitions are known
to occur in a mesoscopic superconductor as $H$ is
changed,~\cite{BaelPeet,SchPeet,SchPeetDeo,DeoSchPeet,Bonca} but
the details are quite different, because different parameter
($\kappa$) regimes and sample geometries (cylinder vs. film) are
studied.

\begin{figure}[htb]
\centerline{\epsfig{figure=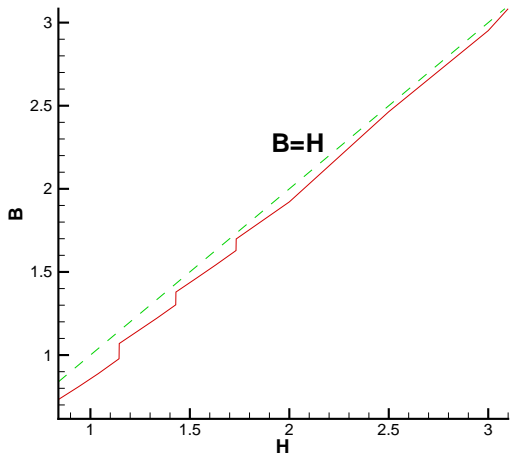,width=8cm}} \vspace{.2cm}
\caption{plotted is $B$ as a function of $H$.}
\end{figure}

Phase transition is also evident in the $n = 12$ case, where the
vortex configuration shows a sudden change in arrangement even
with the same number of vortices for a slight increase of the
applied field from $H = 1.454$ to $H = 1.455$.

In the $B$ vs $H$ plot, $B$ is much lower than $H$ when the number
of vortices is small, but as the vortices increase, the curve
approaches the $B = H$ curve asymptotically. So for the given
sample, the number of vortices and $B$ increase as the applied
magnetic field $H$ increases. This reflects the fact that the
vortices are a form of magnetic penetration.

\section {Time sequence showing vortex entry dynamics}

Fig. 4 shows plots of Cooper pair density for $H = 1.145$ in time.
The number of vortices is 8. The perfect symmetry in the sample
geometry dominates the transient process, but in the middle of the
process the whole configuration makes a rotation to rearrange
itself into a new configuration. (Note that time advances from
3000 to 17500 between the 8th and 9th frames.) The final result is
still a square-symmetric configuration. We note that during the
rotation process the vortex configuration loses some mirror
symmetries of the sample but still preserves the $90^{\circ}$
rotation symmetry. So these transient states do not possess the
full symmetry of the sample. We think that this is possible
because our numerical method has very weekly broken the sample
symmetry. That is, we think that the state just before the
rotation is a metastable state only within the subspace of
configurations preserving the full symmetry of the sample. Thus,
in the actual physical situation when the sample has perfect
symmetry and the temperature is sufficiently low this rotation may
take a very long time to take place. For samples with imperfect
symmetry this relaxation time may be shorter. Since this is a
symmetry-induced qualitative property of the vortex-entry dynamics
in a mesoscopic superconductor, we believe its general validity is
independent of the fact that we have obtained it by solving a
simplified set of TDGL equations which are not truly physical.

\begin{figure}
\centerline{\epsfig{figure=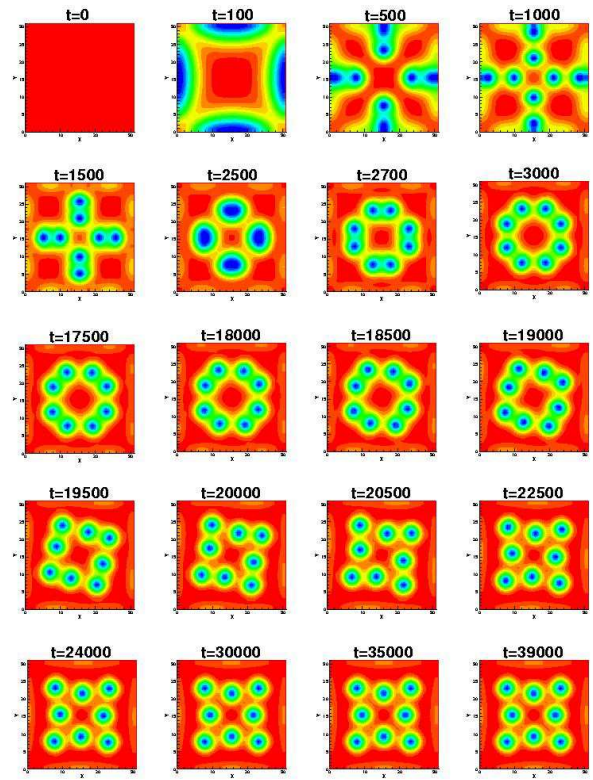,width=8cm}} \vspace{.2cm}
\caption{Time sequence of vortex-entry dynamics for $H = 1.145$.}
\end{figure}

\section {Steady states with reduced symmetry and the equilibrium
state}

The previous sections present solutions for a mesoscopic type-II
superconducting cylinder with initially no vortex inside the
system. The validity of such solutions also requires a perfectly
square sample without any defect at the boundary, and temperature
sufficiently low, so that thermal fluctuations are too weak to
help the system find lower-energy configurations of reduced
symmetry. This is an ideal condition, producing only solutions
consistent with the sample symmetry, and, even during the
transient, the system is bound to this symmetry (except in rare
cases when the transient solutions can keep only fourfold rotation
symmetry but not mirror symmetries
--- See Fig. 4). In principle, one can reproduce this ideal system
in a laboratory with special care.

In real situations there most-likely exist some small defects or
perturbations at the boundary. Then vortices can enter the system
asymmetrically to produce steady-state configurations with reduced
symmetry of lower total Gibbs energy than any symmetric solution.
A strong enough thermal fluctuation could also change the vortex
number and rearrange the vortices to such a configuration. In
previous work, to take into account these perturbations, a random
fluctuation term was added to the governing equation.~\cite{Kato}
This term breaks the symmetry governing the equations, energizing
the system to jump out of the local minima in energy, and over the
energy barrier.~\cite{Tinkham} But this increases the computing
time greatly. (This method is basically what is called ``simulated
annealing''~\cite{SimuAnneal,Doria} in computer science.)

As an alternative approach, we employ perturbed initial conditions
instead of the perfectly superconducting initial condition, and is
similar to Peeters et al.~\cite{SchPeetDeo,BaelPeet} but with a
new idea introduced to make the numerical scheme much more
efficient:

We have first used randomly perturbed initial conditions. They can
indeed lead to final steady-state solutions with reduced symmetry
and lower energies. But we find this way is very inefficient for
finding the equilibrium state at any given $H$. We have also tried
to use a lower-symmetry configuration from such a calculation as
the initial condition for a new $H$ value, but find that the
vortex number can often be trapped in an uncontrollable
non-equilibrium value due to the existence of surface energy
barriers against vortex entry or exit. So this method of adopting
an existing solution as the initial condition can not be reliably
used to find the true equilibrium state in a given system and
field. (Peeters et al. changes the field in small steps to avoid
this difficulty~\cite{Peeters}, but we believe the procedure to be
unnecessarily tedious.)

To obtain the true equilibrium vortex configuration at any given
magnetic field without employing some simulated annealing method,
we have devised a systematic approach to generate initial states
with given numbers of vortices at random distributions. It is
through an analytic expression as follows: First, for one vortex
at the origin in circular coordinates $(r,\theta)$, we use the
widely-known approximate expression~\cite{Schmid,Hu1972,Clem}
\begin{equation}
\Psi(r,\theta) = \frac{re^{i\theta}}{\sqrt{r^2 + \kappa^{-2}}}\,.
\end{equation}
Converting it to Cartesian coordinates, we can move the center of
the vortex to any arbitrary position $(x',y')$ by simply replacing
$(x,y)$ by $(x-x',y-y')$. Denoting this expression as
$\Psi_{x',y'}(x,y)$, an $n$-vortex expression can be simply
constructed as
\begin{equation}
\Psi(x,y) = \Psi_{x_1,y_1}(x,y)\Psi_{x_2,y_2}(x,y)\cdots
\Psi_{x_n,y_n}(x,y)\,.
\end{equation}
This expression obeys the important topological condition that the
phase of $\Psi$ must increase by $2\pi$ when any one vortex center
is circumnavigated. The magnetic field inside the sample does not
obey any topological condition, so it can be simply set equal to
zero for the initial condition. The positions of the vortices can
be generated using random number generators, only if they are
inside the sample. This idea may seem to be simple, but it appears
to have not been employed before. We illustrate below such initial
conditions to obtain steady-state vortex configurations of any
given numbers of vortices $n$. Comparing the total Gibbs energies
of such solutions of different $n$, we can then determine the
equilibrium vortex number and configuration. For illustrative
purposes, we consider the case $H = 0.840$. In Fig. 5, the initial
conditions for $\Psi$ with 1 through 8 randomly-placed
(artificial) vortices [the left figures in (a) through (o)], and
the steady-state vortex configurations they evolve to [the right
figures in (a) through (o)], are shown. The corresponding Gibbs
free energies of these steady states are plotted in Fig. 6 as a
function of the vortex number $n$. The minimum-energy
configuration at $n = 5$ is seen to display the square-symmetry of
a five-vortex configuration with a vortex in the center. Although
we have not yet applied this scheme to other field values, the
method we have devised to find the equilibrium vortex
configurations for a given size and shape of the sample and
different values of the external magnetic field should now be
clear.

\begin{figure} %
\centerline{\epsfig{figure=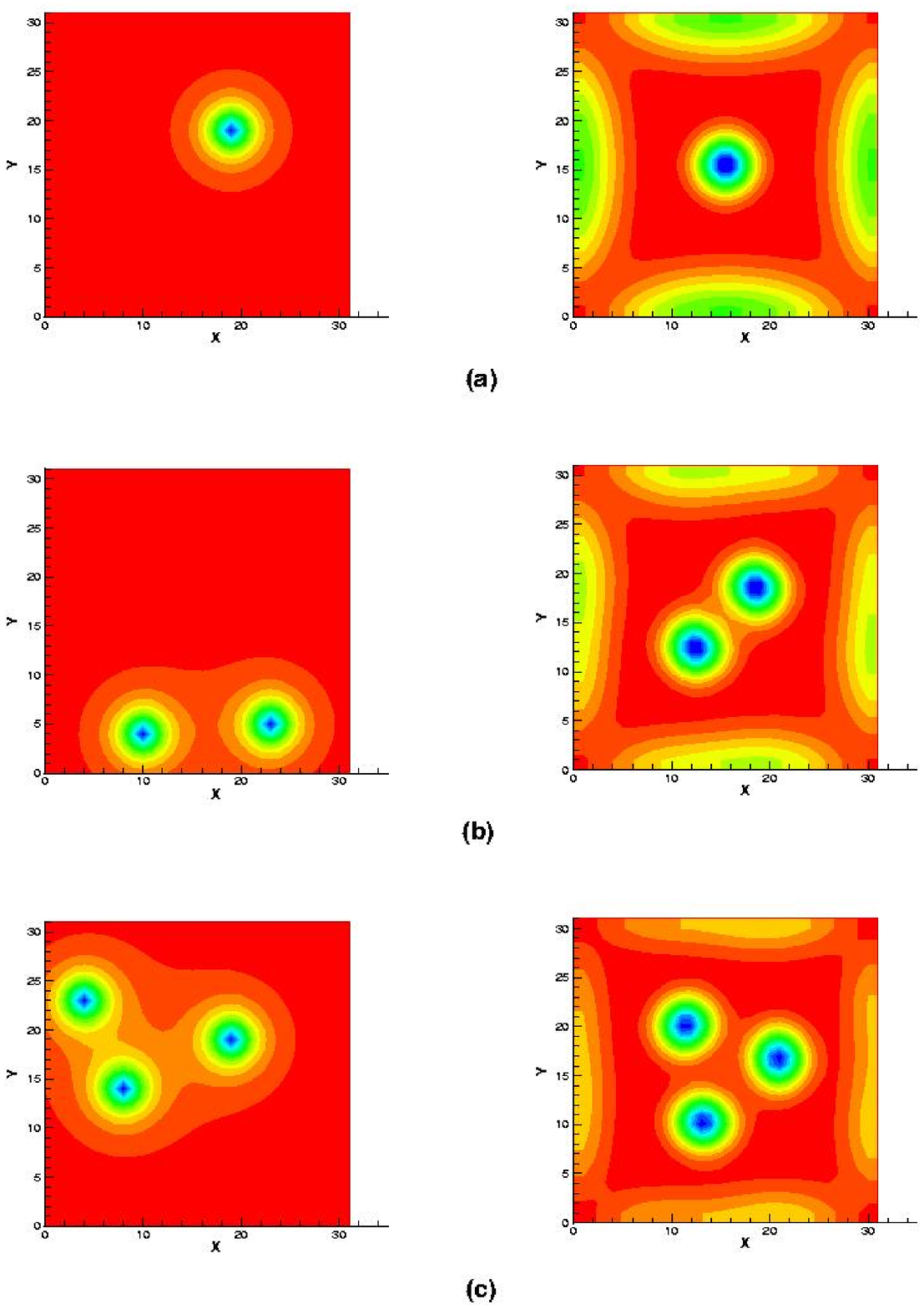,width=8cm}} 
\end{figure} %
\begin{figure} %
\centerline{\epsfig{figure=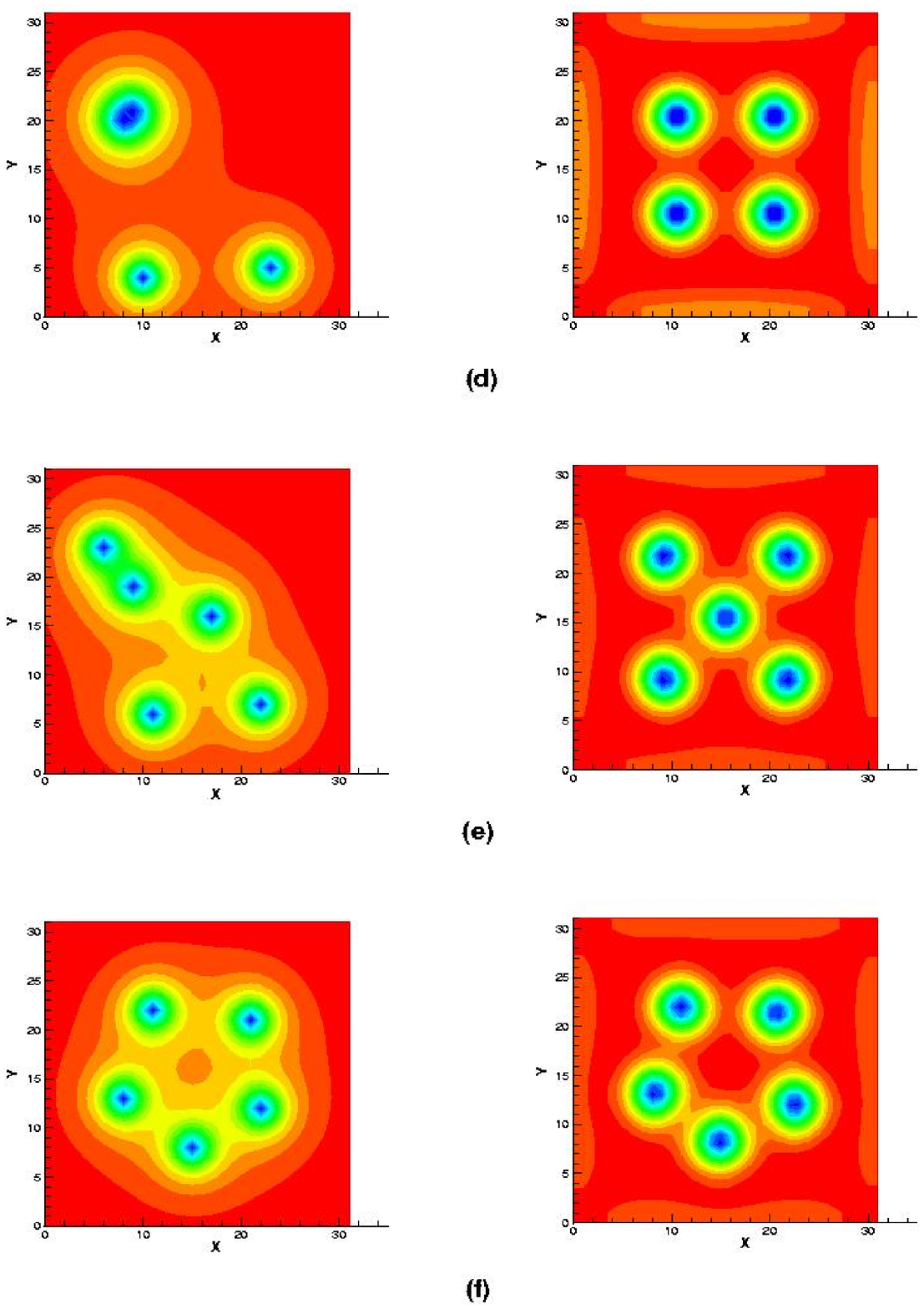,width=8cm}} 
\end{figure} %
\begin{figure} %
\centerline{\epsfig{figure=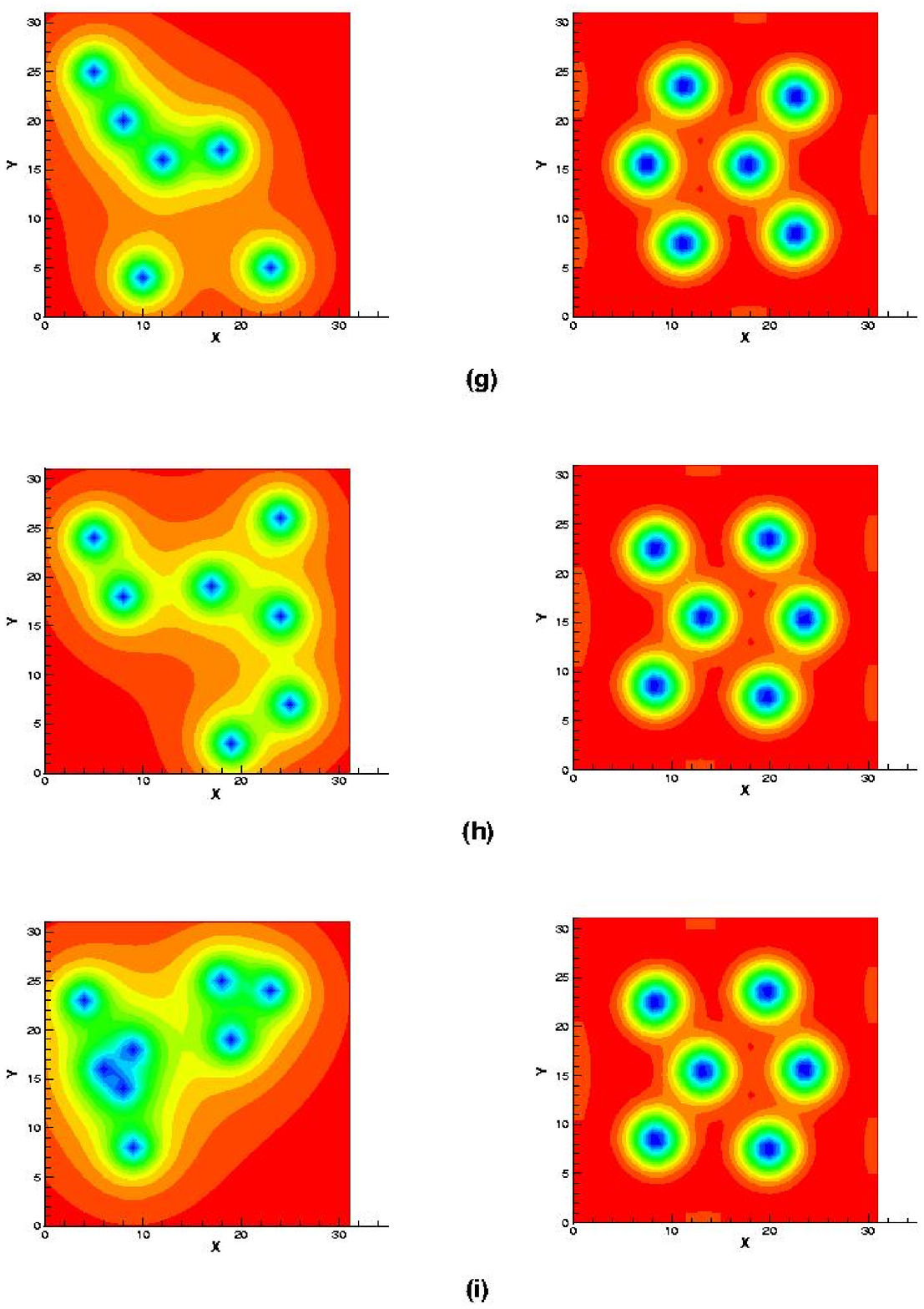,width=8cm}} 
\end{figure} %
\begin{figure} %
\centerline{\epsfig{figure=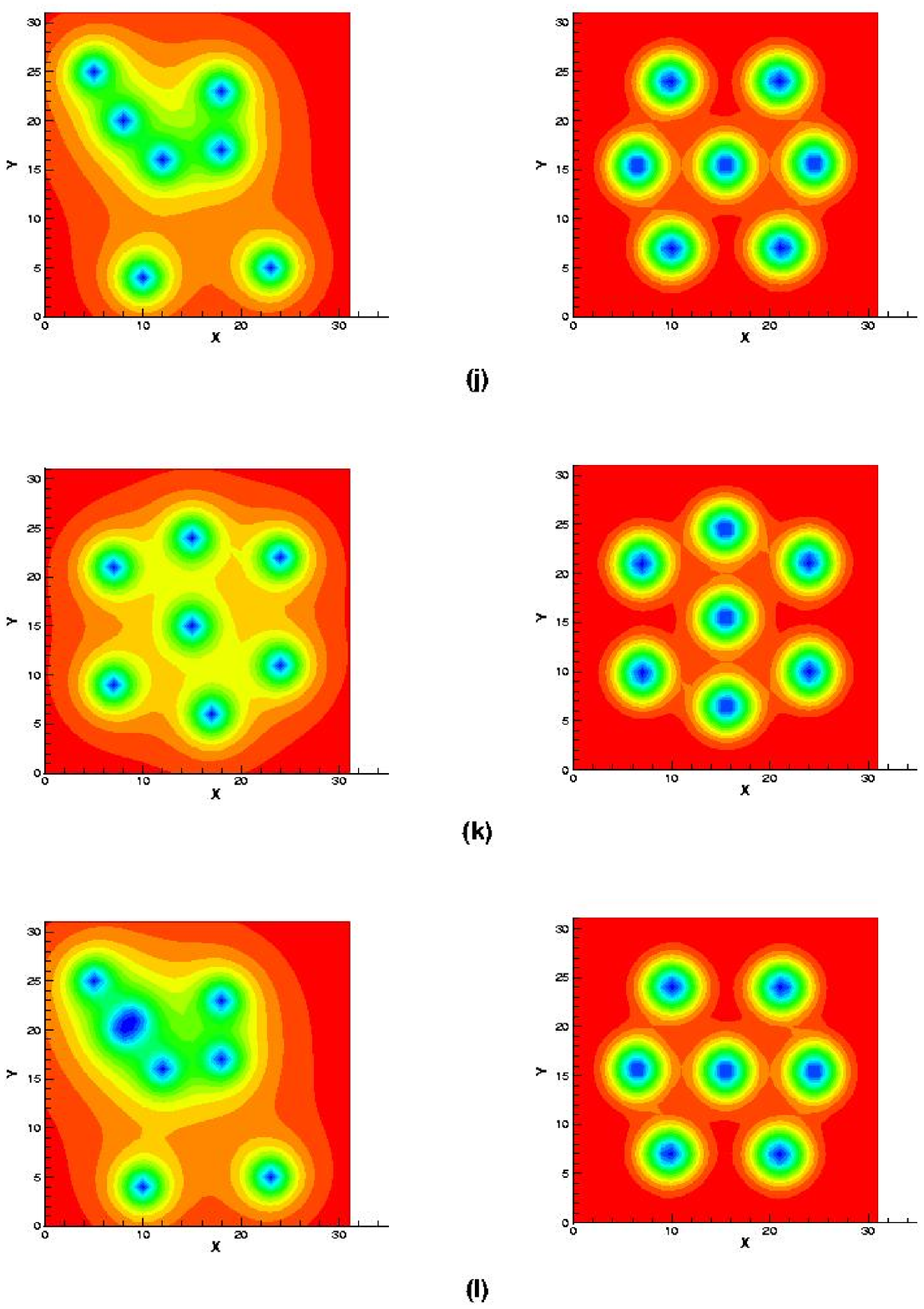,width=8cm}} 
\end{figure} %
\begin{figure} %
\centerline{\epsfig{figure=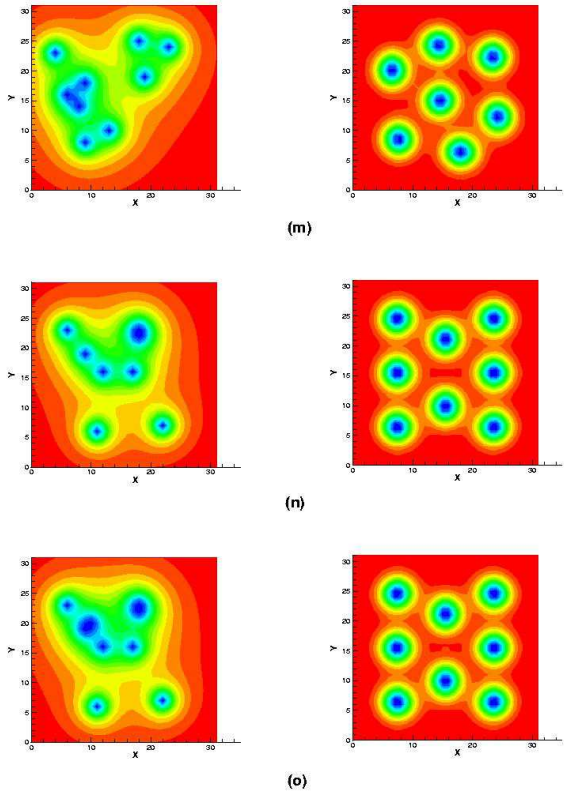,width=8cm}} \vspace{.2cm}
\caption{The initial, random vortex configurations [left figure in
(a) through (o)], and the corresponding steady-state vortex
configurations they evolve to [right figure in (a) through (o)].}
\end{figure} %
\begin{figure} %
\centerline{\epsfig{figure=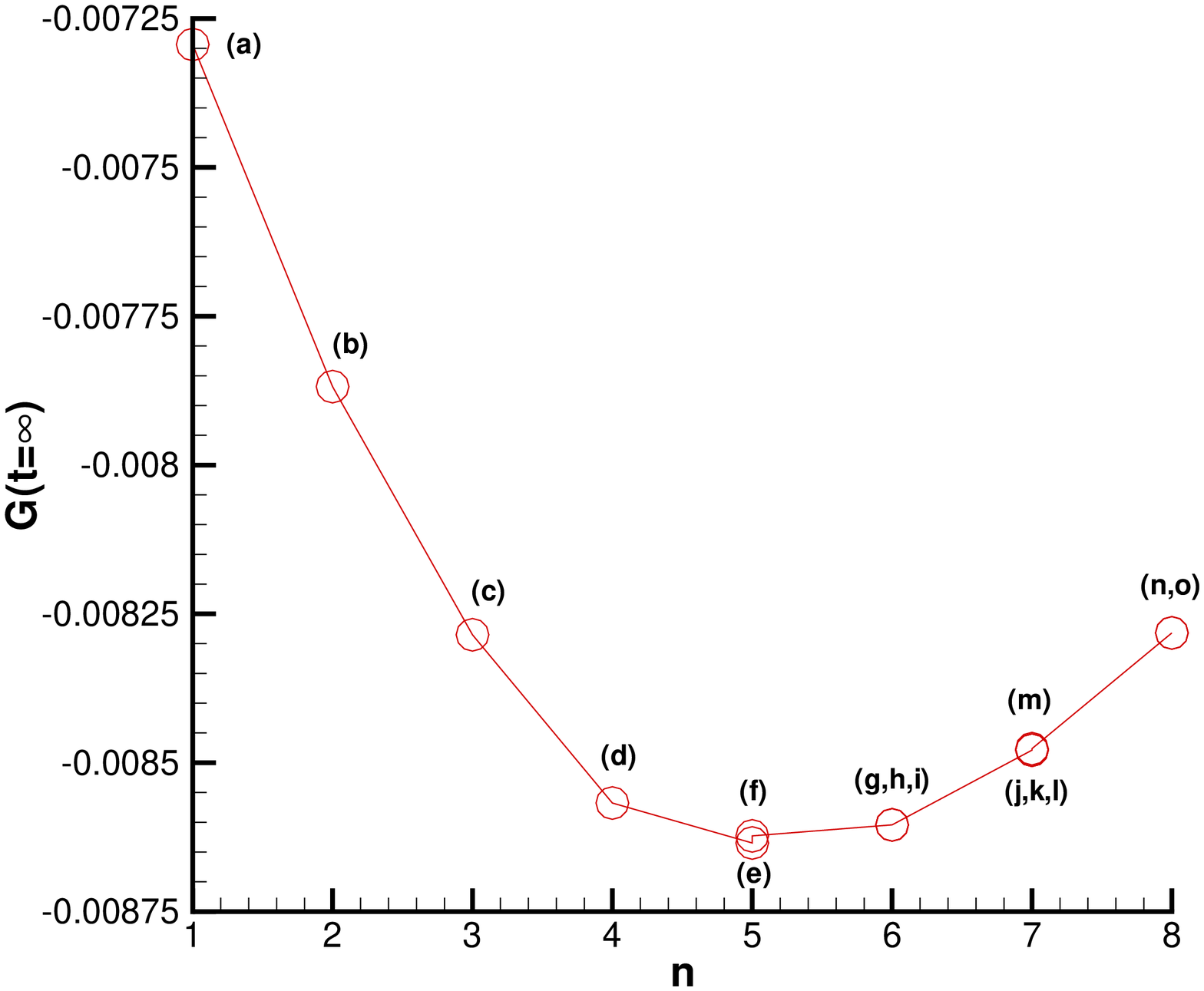,width=8cm}} \vspace{.2cm}
\caption{The steady-state Gibbs free energy for various n.}
\end{figure} %

\section {Summary and Conclusion}

A numerical scheme to study the mixed states in a mesoscopic
type-II superconducting cylinder in a

longitudinal external magnetic field $H$ has been developed. It is
based on solving a set of simplified time-dependent
Ginzburg-Landau equations. We have first applied this scheme to
the case of field penetration into a zero-field cooled sample.
Case studies for various values of the external magnetic field are
presented. Contour plots of the Cooper pair density, and the
induced magnetic field inside the sample, display the magnetic
vortex solution first discovered by Abrikosov, but in a small
sample the vortex arrangement is not simply triangular. Giant
vortices and anti-vortices are not found in this study, unlike
previous studies of type-I mesoscopic thin films. (But at
sufficiently high magnetic field we still expect the system to
favor a single giant vortex at the center as it goes into a
surface superconducting state, but only if the sample is not too
small.) Since we start the solution with a uniformly
superconducting initial condition, and the sample has perfect
square symmetry, both the number of vortices and their
steady-state configurations are governed by the square sample
geometry. Changes in the configuration and the number of vortices
occur as $H$ is varied through first-order configurational phase
transitions, similar to those found earlier, but different in
detail. This phase transition characteristic is confirmed by the
contour plots, and jumps in the values of the induced magnetic
field $B$ at certain discrete $H$ values. A time sequence shows
that the system passes through intermediate configurations, and
remains in some of them for a long time, eventually settling down
to the steady-state configuration, which corresponds to the
lowest-Gibbs-energy configuration consistent with the symmetry
constraints to the vortex number and configuration. One could find
the true equilibrium states, (which would appear in actual
samples, when there are symmetry-breaking surface defects as
vortex-nucleation centers, or when thermal fluctuation is
sufficiently strong to move the system out of metastable states,
but not too strong to melt the vortex lattice,) by adding
additional terms in the equations to simulate thermal
fluctuations, but here we have devised a different approach which
we believe is more efficient. We introduce a way to generate
analytic initial states of prescribed numbers of vortices, but
allow their positions to be random. They evolve to steady-state
vortex arrangements of all possible vortex numbers near the
equilibrium number, from which we can compare total Gibbs energy
to determine the equilibrium vortex number and configuration. In
this way we can avoid the problem of surface and bulk energy
barriers, which can trap the system in non-equilibrium vortex
numbers and configurations --- an undesirable situation which
usually happens if one chooses the initial state randomly without
controlling the vorticity quantum number $L$.

{\bf Acknowledgements} \hspace{.1in} Hu wishes to acknowledge the
support from the Texas Center for Superconductivity and Advanced
Materials at the University of Houston. Andrews acknowledges
support from the Texas A\&M University through the
Telecommunications and Informatics Task Force.


\begin{references}
\bibitem{note3} In this brief survey of literature, we have not included
solutions of linearized GL equations to study the vortex
configurations near the phase boundary, studies of vortex
configurations in the presence of pinning center(s), and studies
based on the London approximation.
\bibitem{note2} As far as we know, previous approaches used random
initial conditions with no control of $L$, so the magnetic field
must be changed in very small steps to obtain speedy convergence.
If so, the present scheme represents an important improvement,
which allows direct implementation at any applied field. Note that
the scheme also applies to the situation when a given distribution
of pinning centers exist in the system or on its surface.
\bibitem{note1} For a larger sample in a sufficiently large
magnetic field, we can occasionally obtain configurations with a
reduced symmetry when solving the equations starting with the
uniform superconducting state, but this is very rare, and very
difficult to control. We think that this is because the numerical
program does not observe the perfect square symmetry. But this
violation of symmetry is a weak one. This way to get an asymmetric
steady state is not reliable, so we do not recommend its use to
find any asymmetric equilibrium state.
\end{references}
\end{document}